\documentclass[twocolumn,aps,prr,superscriptaddress,longbibliography,nofootinbib]{revtex4-2}
\usepackage[T1]{fontenc}
\usepackage[latin9]{inputenc}
\usepackage{color}
\usepackage{babel}
\usepackage{amsmath}
\usepackage{amssymb}
\usepackage{graphicx}
\usepackage[unicode=true,pdfusetitle,
 bookmarks=true,bookmarksnumbered=false,bookmarksopen=false,
 breaklinks=false,pdfborder={0 0 1},backref=false,colorlinks=true]
 {hyperref}
\hypersetup{
 pdfborderstyle=,linkcolor=blue,citecolor=cyan}

\makeatletter

\usepackage{babel}

\makeatother

\begin{document}
\title{Charge-dependent directed flows in heavy-ion collisions by \\ Boltzmann-Maxwell
equations}

\author{Jun-Jie Zhang}

\affiliation{Northwest Institute of Nuclear Technology, Xi'an 710024, China}

\author{Xin-Li Sheng}


\affiliation{Key Laboratory of Quark and Lepton Physics (MOE) and Institute of
Particle Physics, Central China Normal University, Wuhan 430079, China }
\author{Shi Pu}


\affiliation{Department of Modern Physics, University of Science and Technology
of China, Hefei 230026, China}
\author{Jian-Nan Chen}
\affiliation{Northwest Institute of Nuclear Technology, Xi'an 710024, China}
\author{Guo-Liang Peng}
\affiliation{Beijing Institute of Technology, Beijing 100081, China}
\affiliation{Northwest Institute of Nuclear Technology, Xi'an 710024, China}
\author{Jian-Guo Wang}
\affiliation{Northwest Institute of Nuclear Technology, Xi'an 710024, China}
\affiliation{School of Information and Communications Engineering, Xi'an Jiaotong
University, Xi'an 710049, China}
\author{Qun Wang}


\affiliation{Department of Modern Physics, University of Science and Technology
of China, Hefei 230026, China}
\date{\today}

\begin{abstract}
We have calculated the directed flow $v_{1}$ and charge-dependent directed
flow $\Delta v_{1}$ for pions and protons in Au+Au collisions at
$\sqrt{s_{NN}}=200$GeV by solving the coupled Boltzmann-Maxwell equations
self-consistently. Our numerical results show that $v_{1}$ for pions
and protons are all negative in the positive mid rapidity region and have similar behavior  and magnitude.
In contrast we find a quite different behavior in $\Delta v_{1}$
for pions and protons. The difference lies in that $\Delta v_{1}$
for protons mainly comes from pressure gradients of the medium, while
the dominant contribution to $\Delta v_{1}$ for pions is from electromagnetic
fields. Our results indicate that the effect of the electric field
will slightly exceed that of the magnetic and lead to a small negative
slope of $\Delta v_{1}$ for pions.
\end{abstract}
\maketitle


\section{Introduction}

The quantum matter under strong electromagnetic (EM) fields is an
old but still thriving research area in many disciplines of physics.
The strongly coupled quark-gluon plasma (QGP), a new state of matter
governed by Quantum Chromodynamics (QCD), has been produced and extensively
studied in high energy heavy-ion collisions for decades at Relativistic
Heavy Ion Collider (RHIC) of Brookhaven National Lab and at Large
Hadron Collider (LHC) of the European Organization for Nuclear Research
(CERN). 
In the early stage of heavy-ion collisions, extremely strong EM fields
of the order $10^{18}\sim10^{19}$ Gauss are generated \citep{Skokov:2009qp,Bzdak:2011yy,Voronyuk:2011jd},
which leave an imprint on the subsequent evolution of the QGP (for
recent reviews of heavy ion collisions and QGP, e.g., Refs. \citep{Gyulassy:2004zy,Rischke:2003mt,Busza:2018rrf}).
Strong EM fields lead to many novel quantum phenomena such as the
chiral magnetic effect \citep{Kharzeev:2007jp, Fukushima:2008xe}
and the chiral magnetic wave \citep{Burnier:2011bf,Kharzeev:2010gd}
in heavy ion collisions (for recent reviews of these effects, see
Refs. \citep{Kharzeev:2012ph,Kharzeev:2015znc}). 


It requires a self-consistent description of EM fields coupled to
the medium to study these effects. For example, the precise information
about the evolution of EM fields is crucial to extract the CME signals
\citep{Kharzeev:2015znc,Zhao:2019hta,Li:2020dwr,STAR:2021mii} which
has been searched for a decade. The EM fields from spectators can
be well described in previous studies \citep{Voronyuk:2011jd,Toneev:2011aa,Deng:2012pc,Tuchin:2013apa,Pu:2016ayh,Pu:2016bxy,Li:2016tel,Siddique:2021smf}
but not for the parts from the medium produced in collisions, because
it is difficult to describe the medium effects from first principle
with unknown transport properties of the strongly coupled medium and
complicated interaction between EM fields and medium particles. The
fully self-consistent treatment of EM fields and the interacting medium
may help unveil the physics and even puzzles behind these phenomena.
So far as we know, due to great numerical challenges, the exact space-time
evolution of EM fields has not been achieved.

One of such an example is the puzzle related to the directed flow
$v_{1}$ and the charge-dependent directed flow $\Delta v_{1}$ \citep{Gursoy:2014aka}.
The directed flow \citep{Poskanzer:1998yz,Ollitrault:1992bk} is defined
as $v_{1}\equiv\langle\cos(\phi-\Phi_{\text{RP}})\rangle$ and reflects
the collective sideward deflection of particles \citep{Rischke:1995pe,Stoecker:2004qu},
with $\phi$ and $\Phi_{\text{RP}}$ denoting the azimuthal angle
of an outgoing particle and that of the reaction plane respectively.
The charge-dependent directed flow is defined as $\Delta v_{1}^{h}\equiv v_{1}(h^{+})-v_{1}(h^{-})$,
which is the difference between the directed flows of charged particles
and their anti-particles, is expected to be sensitive to the EM field
due to the opposite EM forces exerting on particles with opposite
charges. Currently, both the hydrodynamical and transport models give
a similar pattern of $v_{1}$ which agrees with the experiments. However,
the results of $\Delta v_{1}$ from hydrodynamical models \citep{Gursoy:2018yai,Gursoy:2020jso,Dubla:2020bdz,Inghirami:2019mkc,Song:2017wtw,Chatterjee:2018lsx}
disagree with the measurement of $\Delta v_{1}$ --- the theoretical
results show that both the pion's $\Delta v_{1}^{\pi}$ and the proton's
$\Delta v_{1}^{p}$ in Au+Au collisions at $\sqrt{s_{NN}}=200$ GeV
have negative slopes, while the STAR data for $\Delta v_{1}^{\pi}$
show an almost vanishing slope and those for $\Delta v_{1}^{p}$ have
a positive slope \citep{STAR:2014clz}. The transport models \citep{NA49:2003njx,Pandit:2011np,STAR:2011hyh,Han:2019fce,Oliva:2020doe}
also give consistent results of $\Delta v_{1}$ for hadrons \citep{STAR:2014clz}
at high energies, but it is challenging to include the EM effects
self-consistently in these models. Therefore, the influence of the
electromagnetic fields on the evolution of the system is important\cite{Sheikh:2021rew}.


To reconcile the disagreement, it is essential to perform a fully
self-consistent calculation of the QGP evolution coupled to the Maxwell
equations. Most previous studies either treat EM fields as background
fields without back reactions from medium particles \citep{Gursoy:2018yai,Gursoy:2020jso,Dubla:2020bdz}
or adopt a perturbation method with simplified distributions of EM
fields and QGP \citep{Yan2021}.


In this work, we carry out a fully self-consistent simulation of the
dynamical evolution of the QGP in EM fields by solving the relativistic
Boltzmann equations coupled to the Maxwell equations on Graphics Processing
Units (GPUs)\cite{Zhang:2022PST}. Our algorithm naturally incorporates
all the electromagnetic effects including the Lorentz, Coulomb and
Faraday effects, etc. As a first test of our algorithm, we study the
directed flow $v_{1}$ and its charge-dependent component $\Delta v_{1}^{h}$
for pions and protons, trying to unveil the physics behind the $\Delta v_{1}$
puzzles. With the help of the state of art computing power of GPUs,
we are also able to calculate the evolution of the EM fields in heavy-ion
collisions in a more realistic and precise way, providing a more reliable
baseline for many effects related to EM fields such as the CME effect.


\begin{figure}
\begin{centering}
\includegraphics[scale=0.28]{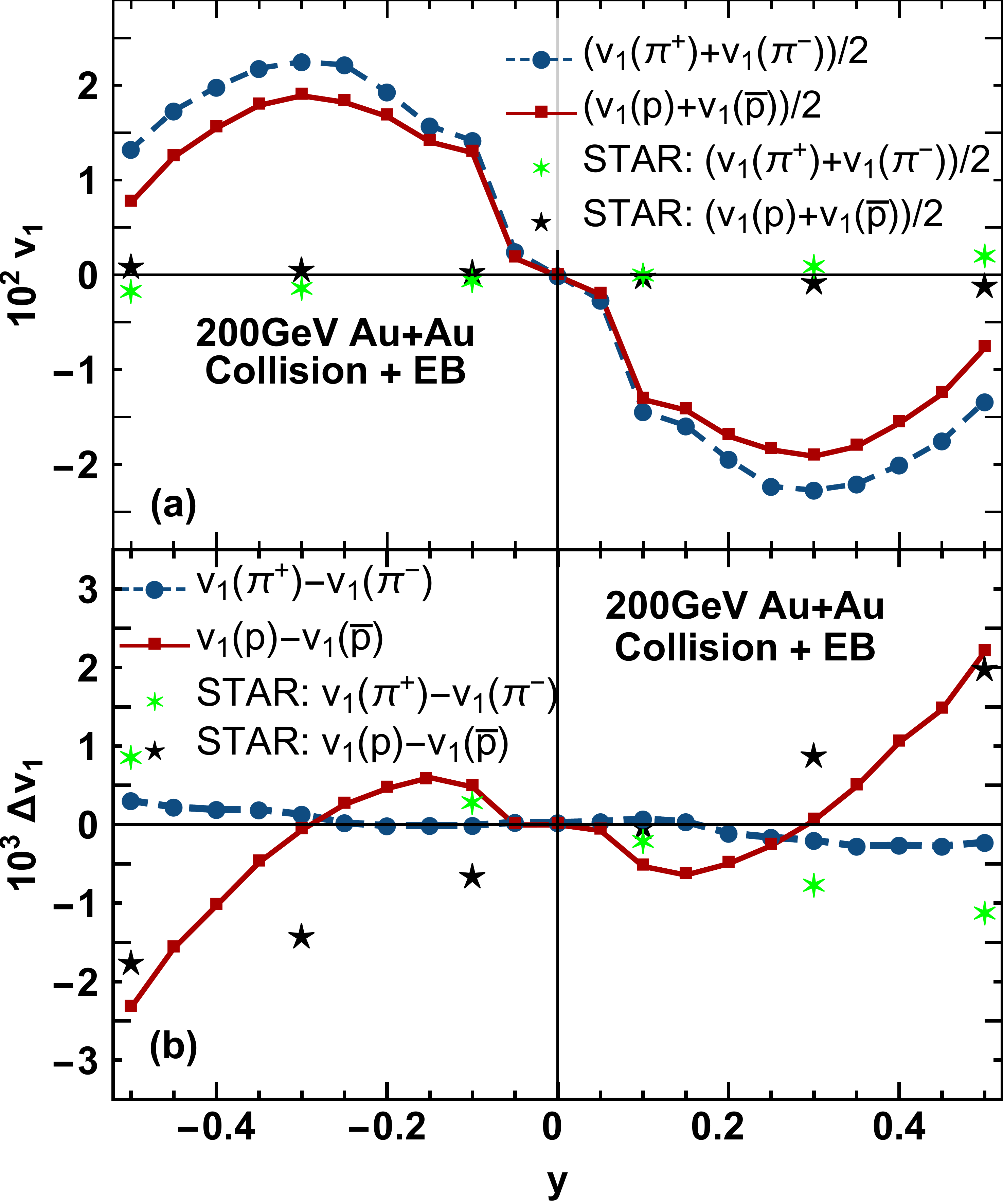} 
\par\end{centering}
\caption{(a) The directed flow $v_{1}$ and (b) charge-dependent directed flow $\Delta v_{1}$
as functions of the rapidity for protons/antiprotons (red solid line)
and $\pi^{\pm}$ (blue dashed line) in Au+Au collisions at 200 GeV. 
The green and black points are derived from  STAR measurements for the $10-40\%$ centrality at $200$ GeV Au+Au collisions \cite{STAR:2014clz}. Although our simulations corresponds to the collisions in $20-30\%$ centrality, it is still observed that the $\Delta v_1$ agrees with the data qualitatively.
The values of the parameters are set to: the saturation time $t=$0.2
fm/c, the impact parameter $b=8$ fm, the factor $r_{q}=0.1$ in Eq.
(\ref{eq:CGC mom}), the strong coupling constant $\alpha_{s}=0.3$,
the range for transverse momenta $p_{T}\in[0.2,\,1.5]$ GeV, and the
distribution functions at $t=$5 fm/c are used. 
}
\label{fig:Calculated-direct-flow} 
\end{figure}


\section{Methods}

The dynamical evolution of the QGP in EM fields is described by the
relativistic Boltzmann equation, 
\begin{equation}
[p^{\mu}\partial_{\mu}+Q_{a}p_{\mu}F^{\mu\nu}\partial_{\nu}^{p}]f_{a}(t,\mathbf{x},\mathbf{p})=\mathcal{C}[f_{a}],\label{eq:BE}
\end{equation}
where $f_{a}$ is the spin and color averaged distribution function
of the parton $a$ with $a=q,\bar{q},g$ for the quark, anti-quark,
and gluon respectively, and $Q_{a}$ denotes its electric charge.
The strong interaction among partons is encoded in the collision term
$\mathcal{C}[f_{a}]$ \footnote{The collision term includes high dimensional integrals which need
more efficient algorithm to calculate. These high dimensional integrals
are calculated by using a powerful numerical integration package ``\textit{ZMCintegral}''
based on GPU developed by some of us \citep{Zhang:2019nhd,Wu:2019tsf}.
For more details, we refer readers to a previous work \citep{Zhang:2019uor}
by some of us. The EM field is also calculated on GPUs using the Jefimenko's
equations \citep{Zhang:2021kpw}.}. In the calculation we consider all 2-to-2 scatterings among u,d,s
quarks, their antiquarks and gluons \citep{Zhang:2019uor}, and the
thermal masses of partons in the matrix elements are chosen to be
$m_{u,d,\bar{u},\bar{d}}=0.3$ GeV, $m_{s,\bar{s}}=0.5$ GeV, and
$m_{g}=0.5$ GeV. The EM field tensor $F^{\mu\nu}$ is determined
by solving the Maxwell equations, 
\begin{eqnarray}
 &  & \partial_{\mu}F_{\alpha\beta}+\partial_{\alpha}F_{\beta\mu}+\partial_{\beta}F_{\mu\alpha}=0,\nonumber \\
 &  & \partial_{\mu}F^{\mu\nu}=j_{\text{ext}}^{\nu}+j_{\text{med}}^{\nu},\label{Maxwell}
\end{eqnarray}
where the source of the EM field has two parts: the external current
$j_{\text{ext}}^{\nu}$ and medium current $j_{\text{med}}^{\nu}$.
The external current $j_{\text{ext}}^{\nu}$ is generated by fast-moving
partons, including spectators and quarks in the rapidity range $|y|>1$.
The dynamical evolution of $j_{\text{ext}}^{\nu}$ is assumed to be
decoupled from the EM field because the trajectories of these fast-moving
particles are hardly influenced by the field. The medium current $j_{\text{med}}^{\nu}$
is from quarks in the mid-rapidity 
\begin{equation}
j_{\text{med}}^{\nu}=\sum_{a=q,\bar{q}}Q_{a}N_{a}\int\frac{d^{3}\mathbf{p}}{(2\pi)^{3}}\frac{p^{\nu}}{E_{a}}f_{a}(t,\mathbf{x},\mathbf{p}).\label{J_med-2}
\end{equation}
where $E_{a}\equiv\sqrt{{\bf p}^{2}+m_{a}^{2}}$ is the energy of
the parton with the mass $m_{a}$ given above, $N_{a}$ is the degeneracy
factor counting the degrees of freedom of the spin and color: $N_{q}=N_{\bar{q}}=6$
for quarks and $N_{g}=16$ for gluons, and the sum runs over all quarks
and antiquarks due to their non-zero electric charges. We see that
$j_{\text{med}}^{\nu}$ leads to a coupling between the Boltzmann
equation (\ref{eq:BE}) and the Maxwell equations (\ref{Maxwell}):
the motion of quarks and antiquarks is influenced by the Coulomb and
Lorentz forces from the EM field, while the EM field is generated
by the motion of charged quarks and antiquarks.



We consider Au+Au collisions at $\sqrt{s_{NN}}=$ 200 GeV. We assume
that one gold nucleus moves along $+z$ direction with its center
located at $x=b/2$ and the other nucleus moves along $-z$ direction
with its center located at $x=-b/2$. We choose the Woods-Saxon distribution
\citep{Vries1987,Shou2015} as the initial \textit{spatial} distribution
for partons in the nucleus. 
The impact parameter is set to $b=8$ fm, corresponding to $20\%-30\%$
centrality approximately. The initial \textit{momentum} distribution
at position $\mathbf{x}$ is inspired by the anisotropic distribution
\citep{Romatschke:2003ms,Romatschke:2004jh} of the Color Glass Condensate
\citep{McLerran:1993ni,McLerran:1993ka,Iancu:2000hn}, 
\begin{equation}
f_{a}(t_{0},\mathbf{x},\mathbf{p})=f_{a}^{(0)}\,r_{q}\,\theta\left(1-\frac{\sqrt{\xi^{2}p_{z}^{\prime2}+\mathbf{p}_{\perp}^{\prime2}}}{Q_{s}}\right),\label{eq:CGC mom}
\end{equation}
where $\mathbf{p}$ and $\mathbf{p}^{\prime}$ are the three-momenta
in the lab and local comoving frame respectively, $Q_{s}$ is the
saturation scale \citep{Iancu:2000hn,Keegan:2016cpi}, $t_{0}\simeq1/Q_{s}$
is the corresponding saturation time \citep{Kurkela:2018vqr,Kurkela:2018wud,Kurkela:2018xxd},
and $\xi$ is the anisotropy parameter \citep{KChurchill2021}. Their
values are chosen to be $Q_{s}=$ 1 GeV, $t_{0}\approx0.2$ fm/c,
and $\xi=1.4$. In the overlapped region of collisions only a fraction
$r_{q}$ of the participants are left in the midrapidity $\left[-1,\,1\right]$.
We fix $r_{q}\simeq0.1$ by making a comparison of the net charge
distribution 
obtained in our simulation with that in the AMPT simulation. 
The coefficients $f_{a}^{(0)}$ for each species of quarks is determined by the corresponding quark number in the overlapped region, i.e., $f_{u}^{(0)}=0.996$,
$f_{d}^{(0)}=1.14$ and $f_{s}=0$. Since there are no anti-quarks initially, $f_{\bar{u},\bar{d},\bar{s}}^{(0)}=0$.
For gluons, we choose $f_{g}^{(0)}\simeq\alpha_{s}^{-1}/r_{q}$ which
is inversely proportional to the coupling constant \citep{Gribov1983}. Note that, in the kinetic approaches, the initial quarks and gluons are treated as quasi-particles. The effect of the sea quarks is included as the thermal  masses of the quarks and gluons. Only after about 0.2 fm/c, which is the reciprocal of the energy scale about $1$ GeV, the soft gluons interact and convert to quark anti-quark pairs. The set-up used in our paper is a typical initial condition in heavy-ion collisions, e.g. also see Ref. \citep{Yan2021,Kurkela:2018vqr,Kurkela:2018wud,Kurkela:2018xxd}.


We emphasize that only $10\%$ of initial partons in the overlapped
region, quantified by $r_{q}\simeq0.1$, contribute to $f_{a}$ and
$j_{\text{med}}^{\nu}$ in the midrapidity region $\left[-1,\,1\right]$.
The remaining $90\%$ of partons are assumed to follow the rapidity
distribution $f^{\pm}(y)=e^{\pm y/2}/{[4\sinh(y_{\text{beam}}/2)]}$
with $1<|y|<y_{\text{beam}}$, where $\pm$ corresponds to the beam
and target direction, respectively. Here $y_{\text{beam}}=5.36$ is
the beam rapidity for collisions at 200 GeV. The motion of these partons
and spectators generates the external current $j_{\text{ext}}^{\nu}$
and thus provides a background EM field for the dynamical evolution
of $f_{a}$.

In the hadronization stage, partons combine into hadrons in each phase
space grid, whose yields agree with experimental data \citep{PHENIX:2003iij,STAR:2003jwm,STAR:2008med}
for the rapidity density $dN/dy$ for $\pi^{\pm}$, $K^{\pm}$, $p$
and $\bar{p}$ at the mid-rapidity $y=0$. 




\section{Results}

\subsection{Negative slope of $v_{1}$.}

\begin{figure}
\begin{centering}
\includegraphics[scale=0.33]{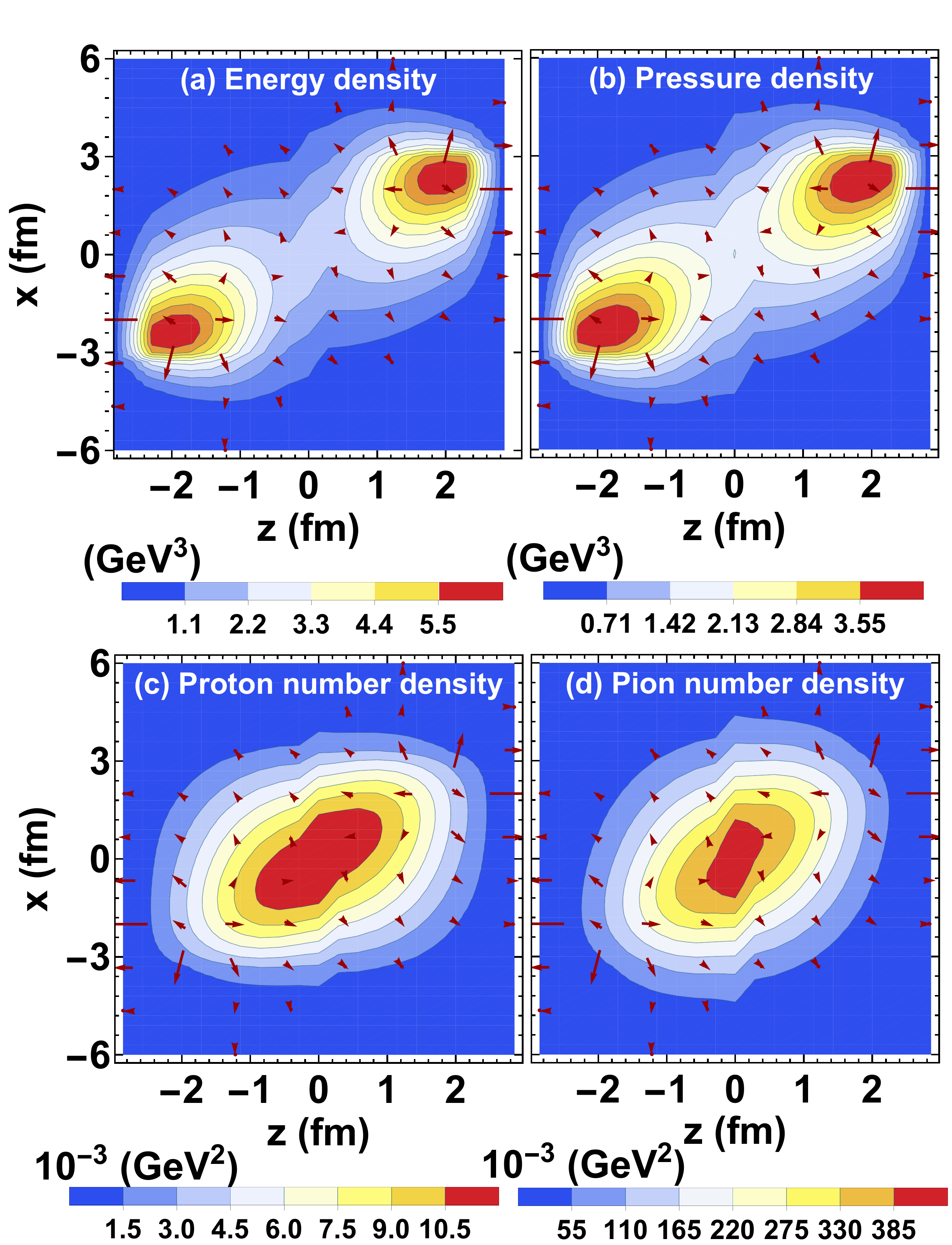} 
\par\end{centering}
\caption{Contour plots for (a) the energy density and (b) the pressure of all
particles in the full rapidity and $p_{T}$ range in the $x-z$ plane.
Contour plots for the number density of (c) protons (without anti-protons)
and (d) pions in the ranges $y\in[-1,1]$ and $p_{T}\in[0.2,\,1.5]$
GeV in the $x-z$ plane. The arrows stand for directions of the pressure
gradients formed by all particles (same for all plots), and the distribution
functions at $t=$2.5 fm/c are used. Other parameters are the same
as in Fig.~\ref{fig:Calculated-direct-flow}.}
\label{fig:pressure} 
\end{figure}

The calculated results for the directed flows as functions of rapidity
for pions and protons in the range $p_{T}\in[0.2,\,1.5]$ GeV are
shown in Fig.~\ref{fig:Calculated-direct-flow}(a). We see that $v_{1}$ for pions and protons have almost the
same magnitude and are positive or negative in $0.4>y>0$ or $0>y>-0.4$ region, respectively. The evolution of the QGP governed by the strong interaction
forms an tilted fireball in the reaction plane as shown by the energy
density and pressure of all particles in the full range of $y$ and
$p_{T}$ in Fig.~\ref{fig:pressure}(a,b). The pressure gradients
lead to an antiflow corresponding to a negative $dv_{1}/dy$ at midrapidity
\citep{Voloshin:2008dg,Brachmann:1999xt,Snellings2000}.

\begin{figure*}
\begin{centering}
\includegraphics[scale=0.28]{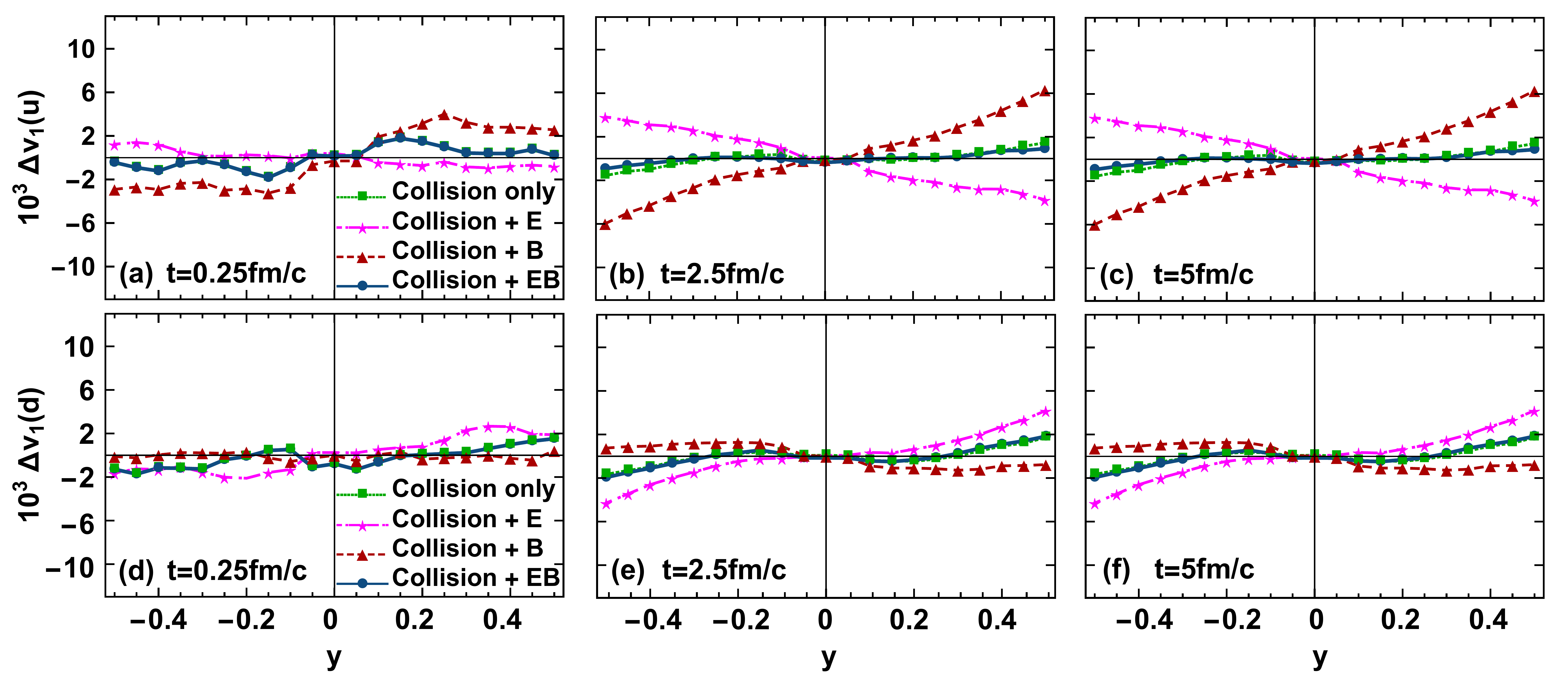} 
\par\end{centering}
\caption{{charge-dependent directed flow $\Delta v_{1}$ for $u$ (upper row) and
$d$ (lower row) quarks at time $t=0.25,\,2.5,\,5.0$ fm/c (left column,
middle column, and right column). The green dotted, magenta dot-dashed,
red dashed, and blue solid curves correspond to cases of collision
only, collision with electric fields, collision with magnetic fields,
and collision with both electric and magnetic fields, respectively.
Other parameters are chosen to be the same as in Fig.~\ref{fig:Calculated-direct-flow}.
} \label{fig:-at-different}}
\end{figure*}

\begin{figure}
\begin{centering}
\includegraphics[scale=0.28]{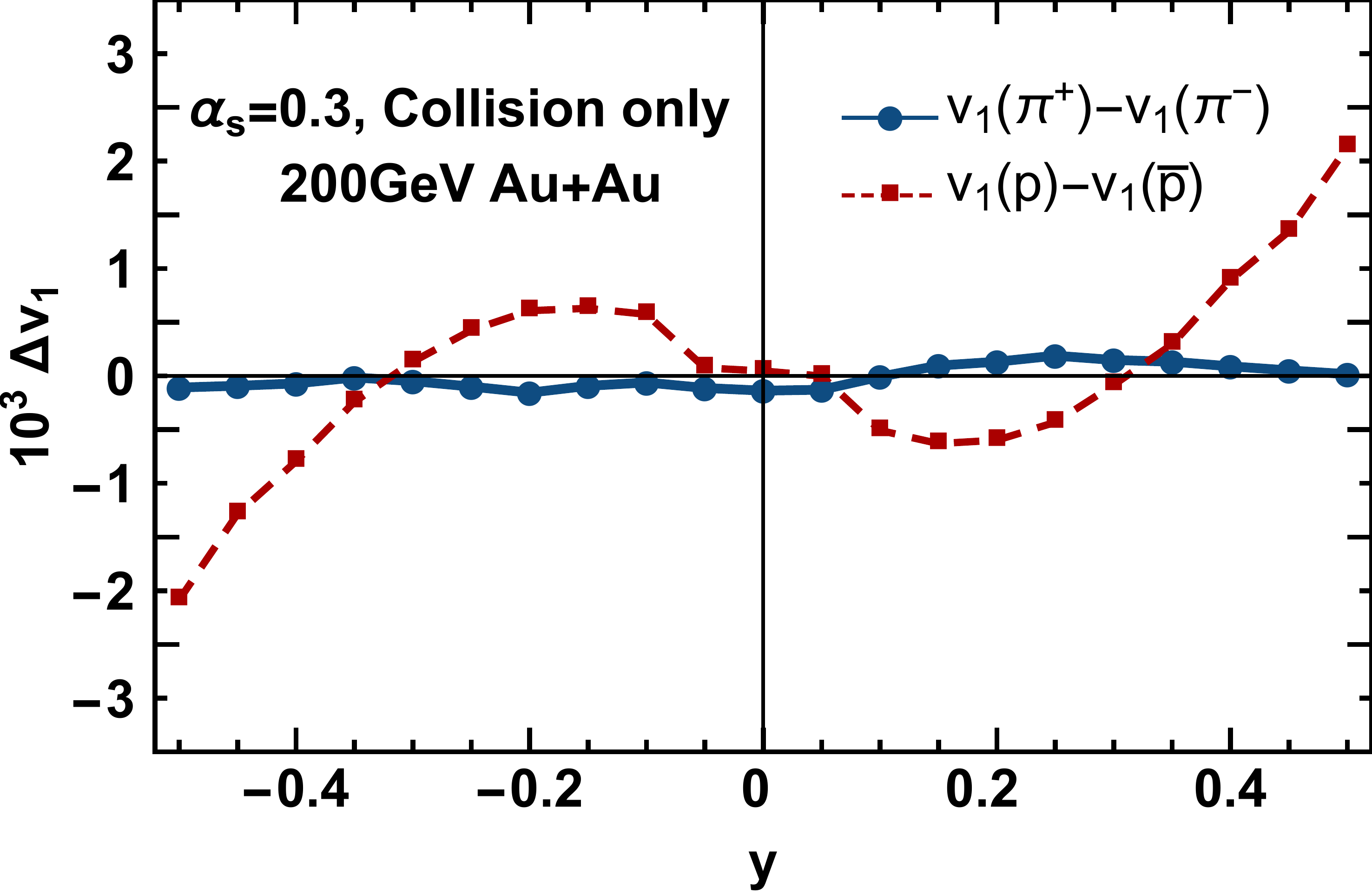} 
\par\end{centering}
\caption{Charge-dependent directed flow $\Delta v_{1}^{\pi}$ and $\Delta v_{1}^{p}$
as functions of rapidity with collisions only (without EM fields).
The parameters are chosen to be the same as in Fig.~\ref{fig:Calculated-direct-flow}.}
\label{fig:dv1C} 
\end{figure}

To understand the generation of the antiflow, we plot in Fig~ \ref{fig:pressure}(c,d)
the contours of the number density for protons and pions in the ranges
$y\in[-1,1]$ and $p_{T}\in[0.2,\,1.5]$ GeV, which are mainly located
in the central region. 
Due to pressure gradients, the protons at forward rapidity, which
are mainly located at $x\simeq0$, $z>0$ region, receive a force
pointing to the right-bottom direction and leading to a negative $v_{1}$.
Similarly, protons at backward rapidity have a positive $v_{1}$.


We also notice the difference between our results and the  STAR data in Fig.~\ref{fig:Calculated-direct-flow}(a). First, our result of $v_{1}$ is more than ten times larger than the experimental data, indicating a larger pressure gradient than the experiments.
We have studied the parameter dependence for $v_1$, such as the initial parton numbers $f_{u}^{(0)}, f_{d}^{(0)}$ and $f_{g}^{(0)}$, the coupling constant $\alpha_{s}$, the size of the spatial grids, the evolution time of the snapshot. 
We find that if more particles are involved in the initial condition (hence larger values for $f_{u}^{(0)}, f_{d}^{(0)}$ or $f_{g}^{(0)}$), a larger pressure gradient can be induced, leading to a larger magnitude of $v_{1}$. If we increase coupling constant $\alpha_{s}$, the system will evolve more like a fluid with stronger collective motions, which narrow the difference between $p_{x}$ and $p_{y}$, thus leading to a smaller magnitude of $v_{1}$.  The value of $v_{1}$ is not sensitive to the size of the spatial grids and the evolution time after  $5$ fm/c. Secondly, we observe that  $|v_1(p)+v_1(\bar{p})|<|v_1(\pi^+)+v_1(\pi^-)|$ in our result, while an opposite behavior is found in the data. Such difference may mainly come from the hadronization model, since in the current study, we consider a simple coalescence hadronization model.
To get a better understanding of $v_1$, a systematical study on the hadronization model in the current framework is required and will be present somewhere else. More discussion on hadronization model at quark level can also be found in Sec. \ref{subsec:dv1_quarks}.

\subsection{Different behaviors of $\Delta v_{1}$ for pions and protons.}

The results for charge-dependent directed flows $\Delta v_{1}$ for pions
and protons are presented in Fig.~\ref{fig:Calculated-direct-flow}(b).
Since the dynamics of all charged quarks are governed by the same
EM fields, a natural expectation is that $\Delta v_{1}$ for pions
and protons as functions of rapidity should be similar, which has
been observed in studies of hydrodynamics incorporating the EM fields
\citep{Gursoy:2018yai,Gursoy:2020jso,Dubla:2020bdz}. However, we
find in our study that $\Delta v_{1}^{p}$ has a positive slope while
$\Delta v_{1}^{\pi}$ has a very small negative slope. The reason that we observe a little negative slope of $\Delta v_{1}^{p}$ in rapidity range [-0.2, 0.2] might due to the over simplified model of the hadronization process. In Fig.~\ref{fig:-at-different}, we do not observe such phenomenon at the quark level.

How to understand such counter-intuitive results of $\Delta v_{1}$
for pions and protons? In fact, the different behaviors of $\Delta v_{1}^{\pi}$
and $\Delta v_{1}^{p}$ come from an interplay of pressure gradients
and EM fields. 


The positive slope for $\Delta v_{1}^{p}$ is mainly attributed to
pressure gradients, similar to the difference between $v_{1}(\pi^{+})+v_{1}(\pi^{-})$
and $v_{1}(p)+v_{1}(\bar{p})$ as shown in Fig.~\ref{fig:Calculated-direct-flow}.
Antiprotons as newly produced particles are more likely to appear
in the region with higher energy densities and therefore larger pressure
gradients as observed in Fig.~\ref{fig:pressure}. Therefore the
negativity for $v_{1}(\bar{p})$ is enhanced relative to $v_{1}(p)$.
Such an effect exists even when the EM fields are switched off. In
Fig.~\ref{fig:dv1C}, we turn off the EM fields and plot $\Delta v_{1}$
caused by collisions only. We observe that $\Delta v_{1}^{\pi}$ almost
vanish but $\Delta v_{1}^{p}$ still have positive slopes. 
On the other hand, the EM fields will influence the evolution of the
QGP and therefore modify the pressure distribution as well as the
number density distribution of hadrons, which finally results in an
amplification of the $\Delta v_{1}^{p}$ slope. Our results for $\Delta v_{1}^{p}$
qualitatively agree with the UrQMD simulation \citep{Bleicher:2000sx,Petersen:2006vm}
and the data of STAR experiment at RHIC \citep{STAR:2014clz}. But
our results are quantitatively smaller than the data because the pressure
induced by $j_{\text{ext}}^{\nu}$ in Eq. (\ref{Maxwell}) is neglected
in this work.

The approximately vanishing $\Delta v_{1}^{\pi}$ in Fig.~\ref{fig:dv1C}(a)
indicates that the splitting between $\pi^{+}$ and $\pi^{-}$ in
the transverse plane is a cumulative result of the EM fields. The
small negative slope of $\Delta v_{1}^{\pi}$ in Fig.~\ref{fig:Calculated-direct-flow}(b)
is consistent with the results from hydrodynamics incorporating the
EM fields \citep{Gursoy:2018yai,Gursoy:2020jso,Dubla:2020bdz}. Unlike
the case of protons and anti-protons, $\pi^{+}$ and $\pi^{-}$ receive
similar contributions from pressure gradients since they have almost
identical spatial distributions.



\subsection{$\Delta v_{1}$ dependence on $p_{T}$ range.}

In Fig. \ref{fig:Calculated-direct-flow} (b), the transverse momentum
is chosen in the range $p_{T}\in[0.2,1.5]$ GeV. If, as discussed
in the above, the behaviors of pions and protons in
$\Delta v_{1}$ are from different mechanisms, the result should
not be sensitive to the momentum range. To support the statement,
we also calculate the dependence of $\Delta v_{1}$ on
transverse momentum ranges. A parameter scan of the $p_{T}$ range in range [x, 1.5] GeV, where $x\in \left\{0.2,0.4,0.6,0.8,1.0,1.2\right\}$, shows that $\Delta v_{1}$
is not sensitive to the choice of transverse momentum ranges. We also observe that particles with $p_{T}$ less than
1 GeV have little contribution to the $\Delta v_{1}$ difference
between pions and protons.

\subsection{$\Delta v_{1}$ for quarks.} \label{subsec:dv1_quarks}

A widely discussed issue for $\Delta v_{1}^{\pi}$ is whether the
contribution from the electric field is more important than that from
the magnetic field \citep{Gursoy:2018yai,Gursoy:2020jso,Dubla:2020bdz}.
To answer this question, we take a closer look at $\Delta v_{1}$
for quarks. 
We show the results of $\Delta v_{1}^{u}$ and $\Delta v_{1}^{d}$
as functions of rapidity in collisions with and without the electric
(\textbf{E}) and magnetic (\textbf{B}) fields in Fig.~\ref{fig:-at-different}
at three different times $t=0.25,\,2.5,\,5.0$ fm/c.

For the case of collision only, different spatial distributions of
$u$ ($d$) and $\bar{u}$ ($\bar{d}$) give positive slopes for both
$\Delta v_{1}^{u}$ and $\Delta v_{1}^{d}$, which leads to the positive
slope of $\Delta v_{1}^{p}$ via hadronization.

\begin{figure}
\begin{centering}
\includegraphics[scale=0.28]{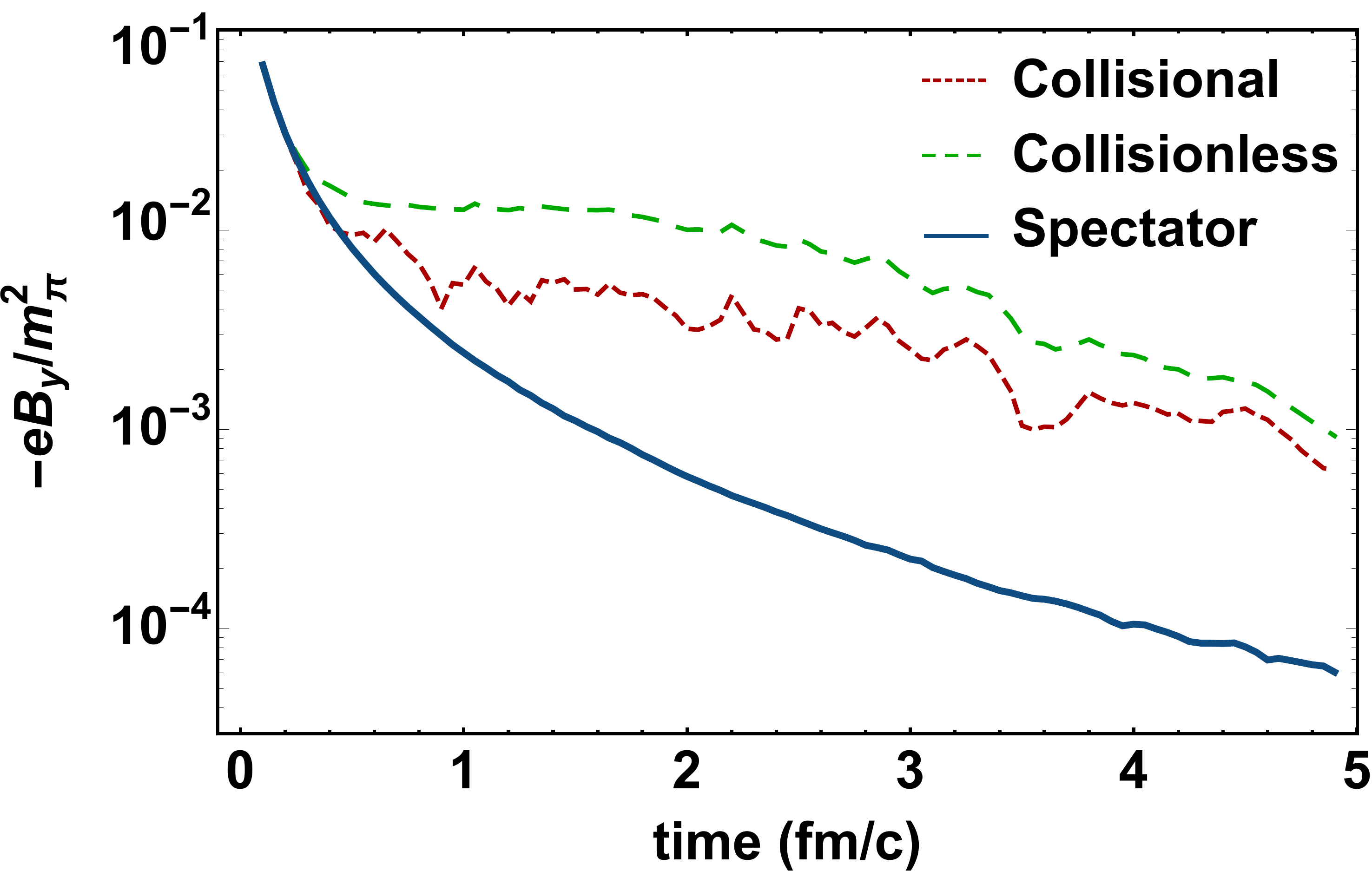}
\par\end{centering}
\caption{{The time evolution of magnetic fields in the central region  of the reaction plane, $(x,y,z)=(0,0,0)$fm.  Other parameters are set to be
the same as in Fig. \ref{fig:Calculated-direct-flow}. } \label{fig:EBVStime}}
\end{figure}

The contributions from electric and magnetic fields to $\Delta v_{1}$
are opposite but in the same magnitude, which agrees with the theoretical
result of Ref. \citep{Gursoy:2018yai}. 
Positively charged particles in forward rapidity are mainly influenced
by the EM field from spectators with $B_{y}<0$ and $E_{x}<0$. Therefore
the magnetic force points to $+x$ direction while the electric force
points to the opposite direction, so two forces partially cancel and
lead to the net effect that is reflected in the difference of $\Delta v_{1}$
between the cases with and without the EM field as shown in Fig. \ref{fig:-at-different}.
We emphasize that the directed flow is a result of accumulation over
time, so the balance of electric and magnetic contributions gradually
changes with time. At an earlier time, e.g., $t=0.25$ fm/c, $\Delta v_{1}$
is almost vanishing which is the result of the cancelation of the
electric and magnetic contributions, while the contribution from the
electric field becomes larger at later time, e.g., $t=5$ fm/c, and
eventually $\Delta v_{1}$ slightly favors the electric contribution.
In Fig. \ref{fig:EBVStime}, we present the evolution of the magnetic field in the central region of the reaction plane, in which the effects form collisions and medium partons can be clearly seen. Similar results can be found in Refs. \cite{Yan2021, Wang:2021oqq}.

Despite the hadronization model used in our simulation, we can understand
the results by a simple sum rule in a naive picture of coalescence
hadronization, i.e., a hadron's $v_{1}$ is approximately equal to
the sum over $v_{1}$ of its constituent quarks \citep{Fries:2003kq,Fries:2008hs,Dunlop:2011cf}.
To this end, we separate the contributions from EM fields and pressure
gradients as $\Delta v_{1}^{u}\simeq2\Delta v_{1}^{\text{EM}}+\Delta v_{1}^{\text{pressure}}$,
and $\Delta v_{1}^{d}\simeq-\Delta v_{1}^{\text{EM}}+\Delta v_{1}^{\text{pressure}}$.
The EM field contribution $\Delta v_{1}^{\text{EM}}$ is proportional
to the quark's charge, while the pressure contribution $\Delta v_{1}^{\text{pressure}}$
is the same for all quarks, as shown in Fig. \ref{fig:-at-different}.
Then following the coalescence sum rule, we have $\Delta v_{1}^{\pi}\simeq\Delta v_{1}^{u}-\Delta v_{1}^{d}\simeq3\,\Delta v_{1}^{\text{EM}}$
and $\Delta v_{1}^{p}\simeq2\Delta v_{1}^{u}+\Delta v_{1}^{d}\simeq3\,\Delta v_{1}^{\text{EM}}+3\Delta v_{1}^{\text{pressure}}$,
with their slopes in agreement with the results in Fig. \ref{fig:Calculated-direct-flow}.
Meanwhile, we also find that the slopes of $\Delta v_{1}$ are insensitive
to the coupling constant $\alpha_{s}$.



\subsection{Time variation of effective conductivity.}

\begin{figure}
\begin{centering}
\includegraphics[scale=0.28]{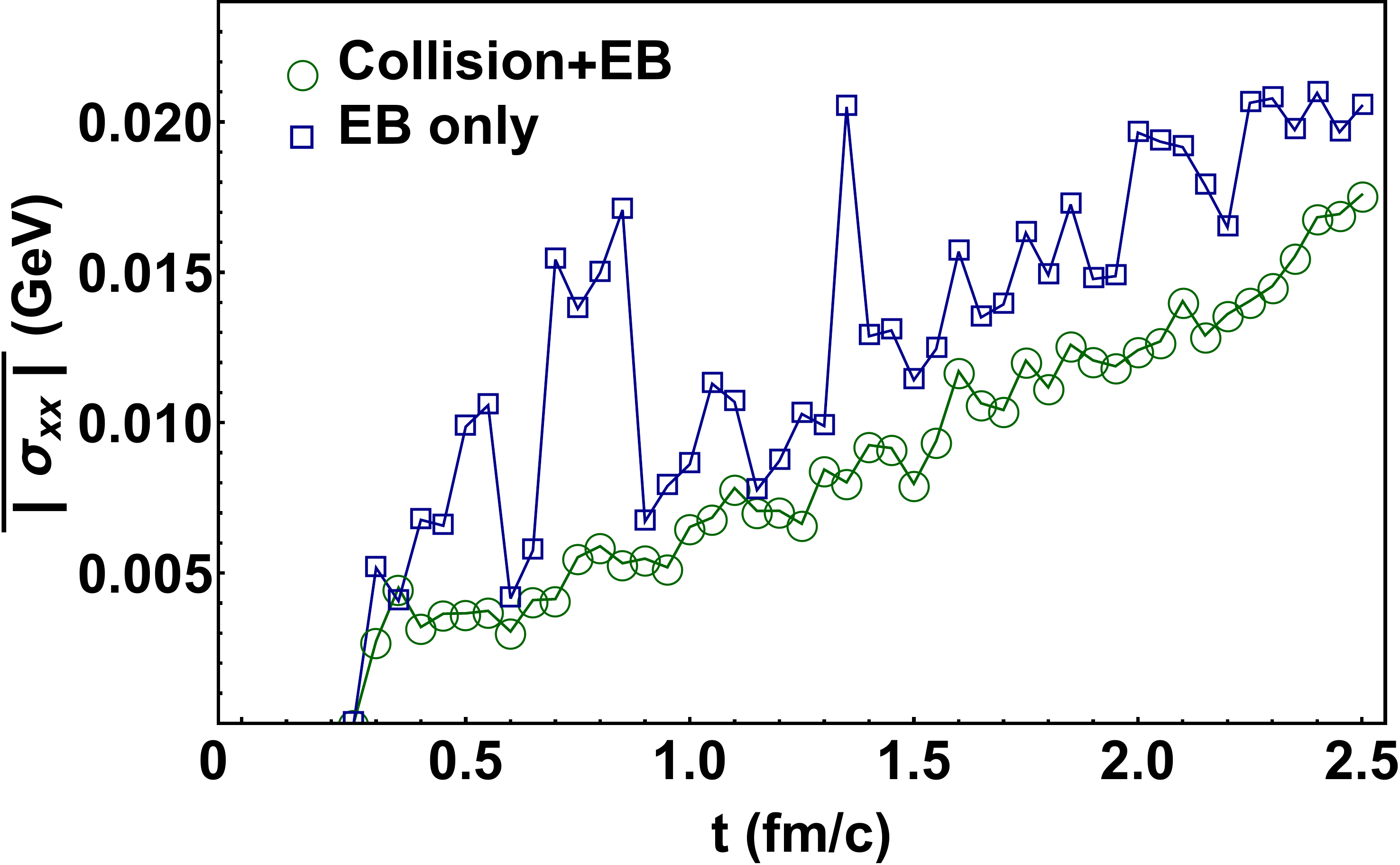}
\includegraphics[scale=0.28]{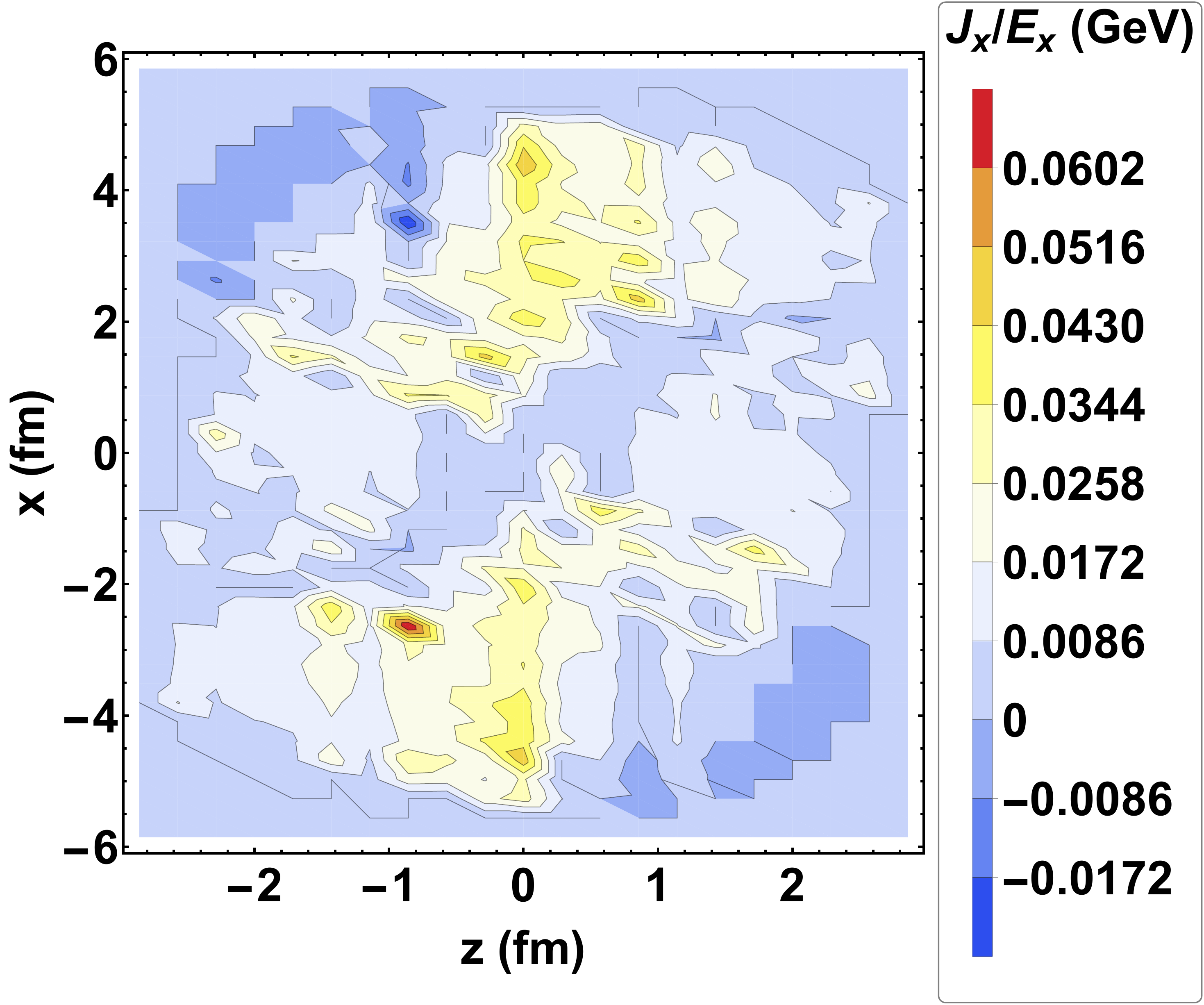}
\par\end{centering}
\caption{{(a) $\overline{|\sigma_{xx}|}$ at different time steps for two
cases: one with collision and one without. (b) Spatial distribution
of $J_{x}/E_{x}$ in the reaction plane. Other parameters are set
the same as in Fig. \ref{fig:Calculated-direct-flow} } \label{fig:(a)--at}}

\end{figure}

An important quantity in the evolution of the quark gluon plasma is
the Ohmic conductivity. Since the system has not reached local thermal
equilibrium the conductivity is a tensor rather than a scalar. To
perceive the presence of the conductivity, we define an effective conductivity
$\sigma_{xx}=J_{x}/E_{x}$ which is a function of
spatial positions. In Fig. \ref{fig:(a)--at} (a), we show the time evolution of the average
absolute value $\overline{|\sigma_{xx}|}=\left[\sum_{i,j}|\sigma_{xx}(x_{i},y_{j})|\right]/n$,
where $n$ denotes the number of grids in reaction plane.
in the reaction plane, which roughly reflects
the amplitude of the Ohmic conductivity. We can see that
when particle collisions are turned on, the conductivity is more
stable with lower magnitude than the collisionless case, consistent with our expectation. In
Fig. \ref{fig:(a)--at} (b) we give the spatial distribution of $\sigma_{xx}$
in the reaction plane at time $2.5$ fm/c, from which we can see that
$\sigma_{xx}$ can be either positive or negative locally.

\section{Discussion and conclusion}

With the help of the state of art parallel computation algorithm,
we are able to calculate the direct flow $v_{1}$ and charge-dependent direct
flow $\Delta v_{1}$ for pions and protons in heavy-ion collisions
by solving the coupled Boltzmann-Maxwell equations for QGP self-consistently.
The collision configuration is set to Au+Au collisions at 200 GeV
and 20-30\% centrality. 

Our numerical results show that $v_{1}$ for pions
and protons are all negative or positive in the $0.4>y>0$ or $0>y>-0.4$ region, respectively and have similar behavior  and magnitude.
The magnitude and behavior of the $v_{1}$ for both protons and pions are different with the experimental data, suggesting that a fine tune of the parameters and a better hadronization model are required.

Our results in the slopes of 
$\Delta v_{1}$ in midrapidity are in a qualitative agreement with
the STAR data.We found that the positive slope of $\Delta v_{1}$ for protons comes
mainly from pressure gradients in the fireball, while the small negative
slope of $\Delta v_{1}$ for pions reflects the contribution from
EM fields over a period of time. The electric and magnetic fields
have opposite contributions to $v_{1}$ and $\Delta v_{1}$ but with
the same magnitude. At a relatively later time, the electric effects
will slightly exceed the magnetic effects, which gives rise to the
small negative slope of $\Delta v_{1}$ for pions. Our results are
insensitive to the values of the coupling constant and can be understood
by a simple sum rule in a naive coalescence picture of hadronization.

To see clear effects from the EM fields, $\Delta v_{1}$ for $D_{0}$
and $\bar{D}_{0}$ mesons may be a better candidate, which needs to
increase the number of momentum grids for heavy quarks. However, restricted
by the GPU resources, our current algorithm does not allow such a
simple extension.


Our calculation can also give a prediction for $v_{1}(\pi^{\pm})$
in low energy collisions. At highest RHIC energy, no significant difference
between $v_{1}(\pi^{+})$ and $v_{1}(\pi^{-})$ has been observed
due to low statistical significance \citep{STAR:2014clz}. In lower
energy collisions, the EM fields will have longer lifetime and therefore
are expected to induce more sideward deflection for charged particles,
i.e. a more negative slope of $\Delta v_{1}^{\pi}$. This qualitatively
agrees with the experimental observation at 7.7, 11.5, and 19.6 GeV
\citep{STAR:2014clz}.



\vspace{1.5em}



\paragraph*{Acknowledgments.}

We would like to thank Hao-Jie Xu for providing us with the charge
rapdity distribution from AMPT to fix the parameters, and thank Umut
G\"{u}rosy, Krishna Rajagopal and Wen-Bin Zhao for helpful discussions.
The work is partly supported by the National Natural Science Foundation of
China (NSFC) under Grant Nos. 12047528, 12075235, 12105227 and 12135011.



\vspace{1.5em}

\bibliography{Ref}


\end{document}